\documentclass[conference]{IEEEtran}
\IEEEoverridecommandlockouts
\usepackage{cite}
\usepackage{float}
\usepackage{amsmath,amssymb,amsfonts}
\usepackage{graphicx, enumitem}
\usepackage{textcomp}
\usepackage[ruled,vlined,linesnumbered]{algorithm2e}
\usepackage[table,dvipsnames]{xcolor}
\usepackage{tikz}
\usepackage{pythonhighlight}
\usepackage{tabularx}
\usepackage{bm}
\usepackage{makecell}
\usepackage{booktabs}
\usepackage{multirow}
\usepackage{url}
\usepackage{subfigure}
\usepackage{tikz}
\usepackage[none]{hyphenat}
\SetKwIF{If}{ElseIf}{Else}{if}{}{else if}{else}{end if}%
 
\setlist[itemize]{leftmargin=4mm}

\newcolumntype{L}[1]{>{\raggedright\let\newline\\\arraybackslash\hspace{0pt}}m{#1}}
\newcolumntype{C}[1]{>{\centering\let\newline\\\arraybackslash\hspace{0pt}}m{#1}}
\newcolumntype{R}[1]{>{\raggedleft\let\newline\\\arraybackslash\hspace{0pt}}m{#1}} 

\def\BibTeX{{\rm B\kern-.05em{\sc i\kern-.025em b}\kern-.08em
    T\kern-.1667em\lower.7ex\hbox{E}\kern-.125emX}}

\begin{document}
	
\title{Solving Drone Routing Problems with Quantum Computing: A Hybrid Approach Combining Quantum Annealing and Gate-Based Paradigms\\
	{}
	\thanks{This research was funded by the European Union, project OASEES (HORIZON-CL4-2022, grant agreement no 101092702), and the by the Federal Ministry for Economic Affairs and Climate Action (German: Bundesministerium für Wirtschaft und Klimaschutz), project Qompiler (grant agreement no: 01MQ22005A). During the preparation of this work the author(s) used Microsoft Copilot in order to improve the language and readability of the manuscript. After using this tool/service, the author(s) reviewed and edited the content as needed and take(s) full responsibility for the content of the publication.}
}

\author{
	\IEEEauthorblockN{Eneko Osaba\IEEEauthorrefmark{2}\IEEEauthorrefmark{1}, 
        Pablo Miranda-Rodriguez\IEEEauthorrefmark{2}, 
        Andreas Oikonomakis\IEEEauthorrefmark{3}, 
        Matic Petri\v{c}\IEEEauthorrefmark{4},\\
        Alejandra Ruiz\IEEEauthorrefmark{2},
        Sebastian Bock\IEEEauthorrefmark{4}, and
	    Michail-Alexandros Kourtis\IEEEauthorrefmark{3}}
	\IEEEauthorblockA{\IEEEauthorrefmark{2}TECNALIA, Basque Research and Technology Alliance (BRTA), 48160 Derio, Spain}
    \IEEEauthorblockA{\IEEEauthorrefmark{3}National Centre for Scientific Research “Demokritos”, Agia Paraskevi, Greece}
    \IEEEauthorblockA{\IEEEauthorrefmark{4}Fraunhofer Institute for Open Communication Systems (FOKUS), Berlin, Germany}
    \IEEEauthorblockA{\IEEEauthorrefmark{1}Corresponding author. Email: eneko.osaba@tecnalia.com}}
\maketitle

\IEEEpubidadjcol	

\begin{abstract}
This paper presents a novel hybrid approach to solving real-world drone routing problems by leveraging the capabilities of quantum computing. The proposed method, coined \textit{Quantum for Drone Routing} (\texttt{Q4DR}), integrates the two most prominent paradigms in the field: quantum gate-based computing, through the \texttt{Eclipse Qrisp} programming language; and quantum annealers, by means of D-Wave System's devices. The algorithm is divided into two different phases: an initial clustering phase executed using a Quantum Approximate Optimization Algorithm (QAOA), and a routing phase employing quantum annealers. The efficacy of \texttt{Q4DR} is demonstrated through three use cases of increasing complexity, each incorporating real-world constraints such as asymmetric costs, forbidden paths, and itinerant charging points. This research contributes to the growing body of work in quantum optimization, showcasing the practical applications of quantum computing in logistics and route planning.
\end{abstract}


\section{Introduction}\label{sec:intro}

Quantum Computing (QC) represents a revolutionary leap in the field of computation, taking advantage of concepts from quantum physics to process information in fundamentally novel ways \cite{nanda2024explorative}. Today, there are high expectations for this field, mainly due to its theoretical potential to solve problems that are currently intractable for classical devices in areas such as cryptography \cite{pirandola2020advances}, drug discovery \cite{zinner2021quantum}, and optimization \cite{abbas2024challenges}. This paper is focused on the last of these areas.

The community is exploring a number of methods in the field of quantum optimization, with Quantum Annealing (QA, \cite{morita2008mathematical}) and variational quantum algorithms such as the Quantum Approximate Optimization Algorithm (QAOA, \cite{farhi2014quantum}) being the most well-known. When combined with appropriate hardware, these methods could offer near- to mid-term advantages, including significant computational speedups, improved solution quality, and reduced energy consumption. Consequently, significant investments and research initiatives are being channeled into developing practical and scalable quantum devices, with the anticipation that they will revolutionize scientific research and industries. In this work, we will explore the use of the two methods mentioned above.

Despite significant advancements in the field, quantum computers remain in their infancy compared to classical computers. Currently, quantum computers struggle to efficiently solve problems due to the limited number of qubits and their inherent instability. Issues such as decoherence, noise, and information loss, especially in the absence of error correction mechanisms, adversely affect computational performance. Additionally, challenges like quantum gate fidelity and gate noise further impede progress. Consequently, we are currently in an era coined \textit{noisy intermediate-scale quantum} (NISQ, \cite{preskill2018quantum}), which is marked by devices' limitations in effectively handling complex problems.

In spite of this seemingly adverse situation, an increasing number of studies are being published in last years around the solving of real-world problems through QC. This growing number of publications certifies that the community is increasingly focusing on exploring the applications of quantum computers. There are several reasons that have contributed to reaching this situation:

\begin{itemize}
    \item \textit{The advances made in the democratization of QC}, understood as the process of making the technology more accessible to a broader spectrum of researchers \cite{seskir2023democratization}. This has greatly facilitated the arrival of newcomers to the field, encouraging them to explore the potential of both quantum gate-based computers and quantum annealers.
    \item \textit{The development of increasingly larger and better-connected devices}. Despite being in the NISQ era, systems as D-Wave's \textit{Advantage\_System}, made up of 5616 qubits organized in a Pegasus topology \cite{boothby2020next}, have contributed to tackling larger and more complex problems.
    \item  \textit{The publication and improvement of frameworks and languages} specifically created to facilitate the design, implementation, and execution of quantum algorithms, with notable examples such as \texttt{Eclipse} \texttt{Qrisp}\footnote{\url{https://qrisp.eu/index.html}}, Qiskit\footnote{https://www.ibm.com/quantum/qiskit}, NVIDIA's CUDA-Q\footnote{https://developer.nvidia.com/cuda-q}, or Google's Cirq\footnote{\url{https://quantumai.google/cirq}}. The existence of this kind of languages contributes to the building of a multidisciplinary community around QC,  facilitating the entry of practitioners from fields such as artificial intelligence or optimization \cite{precup2020experiment}, who may not have extensive knowledge in areas like physics or quantum mechanics \cite{villar2023hybrid}. In this article, we make use of the \texttt{Eclipse} \texttt{Qrisp} framework.
    \item \textit{The implementation of \textit{ready-to-use} hybrid methods that lighten the method-development phase}. A paradigmatic example of this is the D-Wave's portfolio of techniques coined \textit{Hybrid Solver Service} (HSS, \cite{HSS}), which consists of four techniques that target different categories of inputs and problems types. We will provide more details about this portfolio in upcoming sections, as we use three of its methods in this research.
\end{itemize}

With all this, several fields have already benefited from quantum optimization, highlighting cases such as finance \cite{herman2023quantum}, and logistics \cite{phillipson2024quantum}. In this work, we focus on the latter field. More specifically, in this paper we present a hybrid method, coined \textit{Quantum for Drone Routing} (\texttt{Q4DR}), for solving a real-world drone routing problem. The method's main innovation is the use of the two prominent QC paradigms: quantum gate-based computing, through the \texttt{Eclipse} \texttt{Qrisp} programming language; and quantum annealers, through D-Wave's devices. In addition to describing the system developed, we also demonstrate its utility through three use cases, each with different characteristics and increasing complexity.

The remainder of the paper is structured as follows: the next section provides background information on drone route planning and quantum computing programming frameworks. Following this, Section \ref{sec:method} details the characteristics of the implemented method. In Section \ref{sec:exp}, we demonstrate the applicability of the implemented system. Finally, Section \ref{sec:conc} concludes the paper by summarizing the main findings from the experiments and outlining future work.

\section{Background} \label{sec:back}

This section addresses two pivotal concepts to contextualize the current research. Firstly, in Section \ref{sec:drone}, we will provide a concise overview of the state of the art in the intersection of drone routing and QC, emphasizing the unique aspects of our paper. Secondly, in section \ref{sec:frameworks}, we will discuss programming languages and frameworks within the field of QC, with a particular focus on the language utilized in our research, \texttt{Eclipse} \texttt{Qrisp}, and we will underscore its distinctive features.

\subsection{Quantum Computing for the drone mission planning} \label{sec:drone}

As mentioned above, recent years have witnessed a substantial increase in research dedicated to addressing industrial problems through quantum optimization. One specific type of problem that has garnered particular attention is route planning \cite{osaba2022systematic}. Within this subfield, the planning of drone missions has also been a focal point of study. 

In \cite{tarquini2024testing}, for example, the authors address a problem coined as the \textit{Drone Delivery Packing Problem} (DDPP) using the 2000Q Quantum Processing Unit (QPU) of D-Wave. More concretely, the DDPP aims to optimally assign a set of identical available drones to complete a specified set of customer delivery requests. Considering a battery budget for each drone, the planning must be done while adhering to constraints related to time consistency. The maximum size of the problems addressed in that paper does not exceed 12 visiting points.

The authors in \cite{huang2022variational} propose using a QAOA and a Variational Quantum Eigensolver (VQE, \cite{peruzzo2014variational}) to solve a drone collision avoidance problem. Beginning with a set of drones and a number of predefined possible routes, this problem entails optimally selecting the paths for each vehicle to minimize the risk of collision. Factors such as preparation time and early take-off time are considered in this process. Due to the immaturity of the hardware, the tests conducted in this study are executed on very small instances.

In \cite{davies2024quantum}, a study is presented with a focus on solving a real-world problem. More specifically, that work introduces a QUBO formulation that accounts for various constraints, such as the battery consumption of drones and the transportation of goods along the routes. However, due to hardware limitations, the authors only test their proposal with a toy example consisting of four nodes and a single drone. Consequently, the full potential of that work remains to be demonstrated.

Finally, \cite{bar2024quantum} presents a hybrid quantum-inspired metaheuristic for the optimum mission allocation of unmanned vehicles. The main innovation of that paper lies in the planning of three types of unmanned vehicles (aerial, marine, and ground). The maximum number of vehicles considered by instance in the proposed experimentation is six, while the number of missions to be accomplished (points to visit) is 12. Nonetheless, this work falls outside the scope of our research as it presents a method designed to be executed solely on classical computers.

With all this, the work presented in this article introduces the following innovations:

\begin{itemize}
    \item Inspired by studies such as \cite{feld2019hybrid}, \texttt{Q4DR} divides the problem-solving process into two steps: an initial \textit{clustering phase} where the nodes to be visited are divided into two groups, and a \textit{routing phase} where the path taken by each of the drones is calculated. This two-step resolution allows us to tackle problems of a significantly larger size than those described in the aforementioned studies.
    \item Our research stands out by combining both computational paradigms, utilizing a QAOA in the initial \textit{clustering phase} and a quantum annealer in the later \textit{routing phase}.
    \item We consider real-world constraints that add realism to our research: forbidden paths, open routes, and the optimal selection of itinerant recharging points, among others.
\end{itemize}

\subsection{\texttt{Eclipse} \texttt{Qrisp} programming language} \label{sec:frameworks}

Quantum software connects quantum hardware and provides means to utilize it for real-world applications. The majority of current frameworks such as Qiskit, Cirq, Pennylane, TKET, among others, are strongly based on the assembler-like quantum circuit model. In this model, the focus is on the manual construction of quantum circuits by adding quantum gates to the corresponding qubits. This is a feasible approach for small qubit scales; however, for the state-of-the-art systems with more than 100 qubits, the writing of complex algorithms proves to be challenging.

\texttt{Eclipse} \texttt{Qrisp} is a Python-based open-source framework designed to allow high-level programming of quantum computers \cite{seidel2024qrisp}. Its distinctive characteristic compared to the frameworks mentioned above is that it steps away from the conventional assembler-like circuit constructing approaches, and instead introduces \textit{Quantum Variables}, which simplify many programming tasks. This design decision also makes programming in \texttt{Eclipse} \texttt{Qrisp} more familiar to classical programmers by focusing on algorithm development in terms of variables and functions rather than constructing circuits on the lower level. Using the paradigm of \textit{Quantum Variables} hides the qubit management from the user and results in a scalable and modular development process of quantum programs. Another standout feature of \texttt{Eclipse} \texttt{Qrisp} is also its efficient resource management; it provides tools for tracking and optimizing the number of ancilla qubits while also allowing users to benefit other implemented feature like automatic uncomputation \cite{seidel2023uncomputation}.

Finally, installing \texttt{Eclipse} \texttt{Qrisp} provides a collection of pre-implemented algorithms and algorithmic primitives such as, but not limited to, Shor's algorithm, quantum backtracking algorithm, VQE, and QAOA. A comparison in \cite{osaba2024eclipse} showed that \texttt{Eclipse} \texttt{Qrisp}'s QAOA module outperforms two other Qiskit QAOA implementations both in the metric of circuit depth, as well as in the metric of approximation ratio, which justifies the use of this programming language in this research.
	
\section{Method} \label{sec:method}

The hybrid system proposed in this paper, \texttt{Q4DR}, aims to solve a real-world oriented drone routing problem. As previously explained, \texttt{Q4DR} will be tested through three use cases of increasing complexity, coined UC1, UC2 and UC3. In a nutshell, the problem consists of a set of $N$ visiting points, each of which must be visited exactly once by any of the drones in the fleet, with the objective of minimizing the total cost of the paths built. Furthermore, the drone fleet consists of two vehicles, which must be used mandatorily, meaning that the number of routes to be calculated is two. The costs  $c_{ij}$ associated with traveling from one point $i$ to another point $j$ are asymmetric, i.e., $c_{ij} \neq c_{ji}$. Additionally, certain paths are prohibited, such that $c_{ij} = \infty$ or $c_{ji} = \infty$.

In UC1, both routes must start and end at a single depot, while in UC2 and UC3, the drones start from depots located at different points on the map. Additionally, in UC2, each route must also end at the same point where it started, completing a Hamiltonian cycle. Finally, in UC3, there is a set of charging points scattered across the map, so each route must end at one of these points. It is important to highlight that each drone is required to visit exactly one of these charging points, which should be selected in a manner that minimizes the total distance of the route.

\begin{figure}[t]
    \centering
    \includegraphics[width=1.0\linewidth]{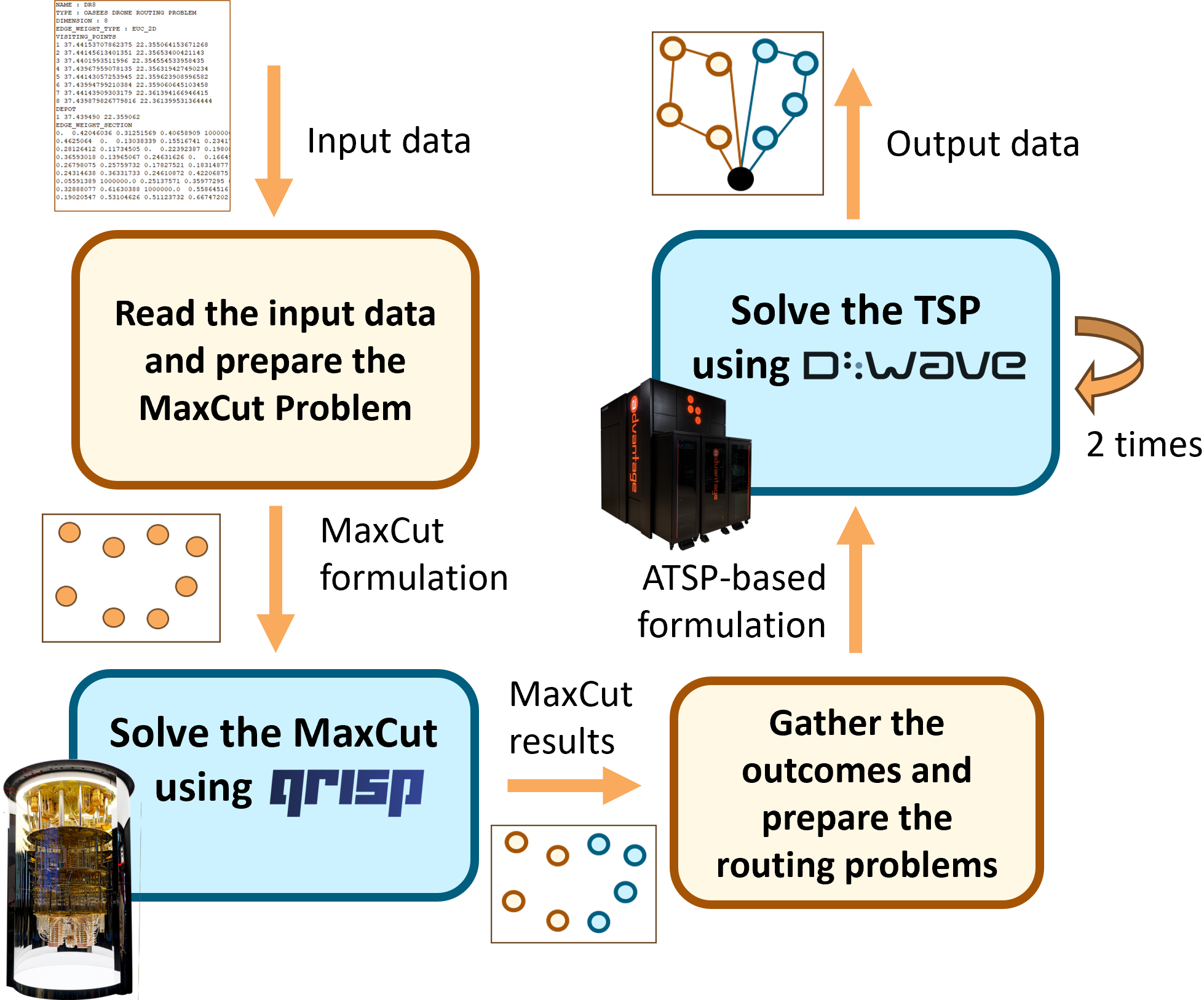}
    \caption{Workflow of the implemented hybrid solver.}
    \label{fig:scheme}
\end{figure}

The workflow of \texttt{Q4DR} is shown in Fig.~\ref{fig:scheme}. This figure illustrates the resolution of the problem in two distinct phases:

\begin{itemize} 
    \item \textit{Clustering phase}, which involves solving the Maximum Cut problem (MaxCut) and is managed by the QAOA module of \texttt{Eclipse Qrisp}. 
    \item \textit{Routing phase}, which utilizes the Asymmetric Traveling Salesman Problem (ATSP) as the basis for the calculations, and employs D-Wave's quantum annealers.
\end{itemize}

The subsequent subsections provide a detailed description of the inner workings of each phase.

\subsection{Clustering phase} \label{sec:maxcut}

The main goal of this phase is to partition the entire graph into two clusters using the MaxCut formulation. This partitioning divides the graph into two regions such that the distances (weights of the edges) between them are maximized. Each node of the graph represents a visiting point on a map, consisting of latitude and longitude coordinates, with the distances represented as the weights of the edges.

Before presenting the details on how \texttt{Eclipse Qrisp} facilitates the implementation of a QAOA algorithm, a brief introduction on how MaxCut can be formulated in QAOA terms follows. In a nutshell, MaxCut is a combinatorial optimization problem where the goal is to partition the vertices of a graph \( G = (V, E) \) into two subsets such that the total weight of edges between the subsets is maximized. The MaxCut problem is typically expressed as the following cost function that needs to be maximized
\[
f(x) = \sum_{(u, v) \in E} \frac{(1 - x_u x_v)}{2} w_{uv},
\]
where:
\begin{itemize}
    \item \( x_u, x_v \in \{-1, 1\} \) represent the subset assignments of vertices \( u \) and \( v \) (\( x_u = 1 \) for one subset, \( x_u = -1 \) for the other),
    \item \( w_{uv} \) is the weight of the edge between vertices \( u \) and \( v \).
\end{itemize}

The following components are the fundamental elements of any QAOA implementation, with specific applications to the MaxCut problem:

\begin{itemize}
    \item \textit{Classical cost function}: the function $f(x)$, where the weight \(w_{uv}\) of the edge between the vertices \( u \) and \( v \) contributes to the overall cost if it belongs to the same partition for a given cut. The cost hamiltonian can be constructed as follows:
    \[
    H_C = \sum_{\langle i, j \rangle \in E} w_{ij} \frac{1}{2}(1 - Z_i Z_j)
    \]
    
    \item \textit{Cost operator or phase separator}: it applies a phase to each computational basis state based on its cost function value: \[
    U_C(\gamma) = e^{-i \gamma H_C} , \gamma \in \{0,2\pi\}
    \]
    
    \item \textit{Mixer operator}: which drives transitions between different states\[
    U_M(\beta) = e^{-i \beta H_M}, \beta \in \{0,\pi\}
    \]
    \item \textit{Initial state}: which usually is a uniform superposition: 
    \[
    | \psi_0 \rangle = \frac{1}{\sqrt{2^n}} \sum_{z \in \{0, 1\}^n} | z \rangle
    \]
\end{itemize}
Using all these fundamentals, a quantum circuit is constructed where the pair of \(U_C\) and \(U_B\) operators form a QAOA layer. After the application of a layer, an angle-dependent state can be defined:
{\small
\[
|\psi_p \rangle = |\gamma,\beta \rangle = U_M(B,\beta_p)U_C(C,\gamma_p)...U_M(B,\beta_1)U_C(C,\gamma_1)| \psi_0 \rangle.
\]
}

Finally the main objective of the QAOA is to find the optimal \(\gamma_p,\beta_p\) that minimizes the expected value of the cost function using classical optimizers \cite{choi2019tutorial}.

Implementing a QAOA requires several essential steps and components. \texttt{Eclipse Qrisp} provides a user-friendly and feature-rich interface that streamlines this process, enabling the implementation of QAOA with just a few lines of Python code. Additional libraries, such as \texttt{networkx} for graph processing and \texttt{geopy} for calculating distances between nodes, can be easily integrated to enhance functionality.

As explained before, this \textit{clustering phase} expects a set of latitude and longitude coordinates as input, constructing a weighted graph based on the distances between all nodes. QAOA is then applied to this graph using \texttt{Eclipse Qrisp}, simplifying every component of the implementation while maintaining efficiency and adaptability. Below, we present the key aspects of the implementation, supported by various excerpts from the developed code.


\begin{itemize}
\item \textit{Input and Graph creation}: The coordinates are represented as a simple array of tuples. This information is taken from the file introduced as input (as seen in Fig.~\ref{fig:scheme}).
\end{itemize}


\begin{python}
import networkx as nx
import geopy
coordinates = [
    (Latitude1,Longitude1),
    ...,
    (LatitudeN,LongitudeN])
G = nx.Graph()
....# graph creation
\end{python}
\begin{itemize}
\item \textit{Classical cost function}: Two different Python methods are implemented for this purpose.  First, \texttt{maxcut\_cost\_funct}, which expects a measurement results in order to calculate the energy of the hamiltonian based on the second method, \texttt{maxcut\_obj}, which is called internally.
\end{itemize}
%
\begin{python}
def maxcut_obj(x):
    cut = 0
    for i, j, data in G.edges(data=True):      
        if x[i] != x[j]:
            cut -= data['weight']
    return cut

def maxcut_cost_funct(meas_res):
    energy = 0
    for meas, p in meas_res.items():
        obj_for_meas = maxcut_obj(meas)
        energy += obj_for_meas * p
    return energy
\end{python}
\begin{itemize}
\item \textit{Cost operator}: \texttt{Eclipse Qrisp} provides all the gates needed to construct the quantum circuit. In the case of the weighted MaxCut, the weight of each edge must be included in the circuit. For this purpose, the \textit{QuantumVariable} resource is employed. This \texttt{Eclipse Qrisp} built-in structure is the quantum equivalent of a regular variable in classical programming and, in the MaxCut case, its size is equal to the total number of graph nodes.
\end{itemize}
%

\begin{python}
from qrisp import h, barrier, rz, rx, cx
from qrisp.qaoa import  QuantumVariable

qarg = QuantumVariable(len(G))
def cost_operator(qv, gamma):
    for u, v, data in G.edges(data=True):
        cx(qv[u], qv[v])
        rz(2 * gamma * weight, qv[v])
        cx(qv[u], qv[v])
        barrier(qv)
    return qv
\end{python}

\begin{itemize}
\item \textit{Mixer operator, Initial state and Optimization}: \texttt{Eclipse Qrisp} has a mixer implementation recommended for the MaxCut. Furthermore, the initial state is the superposition as default using Hadamard gates, and, finally, the whole procedure is executed by the core class called \texttt{QAOAProblem}.
\end{itemize}

%
%
%
\begin{python}
from qrisp.qaoa import  QAOAProblem, Rx_mixer

depth = #specify layers
maxcut_instance = QAOAProblem(
    cost_operator, 
    RX_mixer,
    maxcut_cost_funct)

#specify max_iter
res = maxcut_instance.run(qarg, depth, max_iter)
    
res_str = list(res.keys())[0]
print("QAOA solution: ", res_str)
\end{python}

It is noteworthy that the \texttt{depth} variable indicates the number of layers of the QAOA. Additionally, \texttt{max\_iter} represents the number of iterations the classical optimizer will execute (for this purpose, the COBYLA optimizer has been used). Finally, the output of the application is a binary bit-string that encodes the partitioning of the initial graph into two separate sets based on the parameters that approximate the best solution.

\subsection{Routing Phase} \label{sec:routing}

As seen in Fig. \ref{fig:scheme}, the routing phase is executed two times after the results of the \textit{clustering phase} are calculated and processed. This is due to the fact that the drone fleet in this research consists of two units. Additionally, since the vehicles have the same characteristics, both routes are independently calculated following the same process.

On one hand, in the case of UC1 and UC2, the routes are calculated using implementations of the ATSP. On the other hand, for UC3, a specific mathematical formulation has been designed for this research, with the intention of appropriately considering open routes with charging points. Further details will be given in the upcoming Section \ref{sec:UC3}. Finally, it is worth noting that in all use cases, prohibited paths are considered, which is implemented by raising the cost associated with traveling from point $i$ to point $j$ to infinity ($c_{ij} = \infty$).

Focusing our attention on UC1 and UC2, and for research purposes, four different solvers have been implemented to solve the ATSP. "While the first method is a purely quantum approach that utilizes the QPU of D-Wave, the remaining three approaches are hybrid algorithms from D-Wave's HSS portfolio.

With regard to the QPU, the \texttt{Advantage\_system6.4} device has been employed, which is the latest from D-Wave at the time of writing. This computer boasts 5,616 qubits and over 35,000 couplers arranged in a Pegasus topology~\cite{boothby2020next}. Similar to the hybrid solvers described later, the QPU has been accessed via the \texttt{Leap} cloud service\footnote{https://www.dwavesys.com/solutions-and-products/cloud-platform/}, and the standard \textit{forward annealing} procedure has been run. This process comprises the following main steps: \textit{i}) converting QUBO into a graph, \textit{ii}) minor graph embedding, \textit{iii}) QPU initialization, \textit{iv}) annealing, \textit{v}) readout, and \textit{vi}) postprocessing. Further details on this process can be found in \cite{yarkoni2022quantum}.

Regarding HSS, it comprises a portfolio of hybrid heuristics that effectively combine quantum and classical computation to tackle large optimization problems as well as real-world industrial cases. At the time of writing, the HSS includes four approaches. In this article, we focus on three of them, each designed to address specific problem types: the binary quadratic model (BQM) algorithm, \texttt{BQM-Hybrid}, for problems expressed on binary values; the constrained quadratic model (CQM) technique, \texttt{CQM-Hybrid}, which can handle problems defined on binary, integer, and even real values; and the Nonlinear-Program Hybrid Solver, \texttt{NL-Hybrid}, which excels at natively permitting nonlinear (linear, quadratic, and higher-order) inequality and equality restrictions, even when expressed arithmetically.

\begin{figure}[t]
    \centering
    \includegraphics[width=0.9\linewidth]{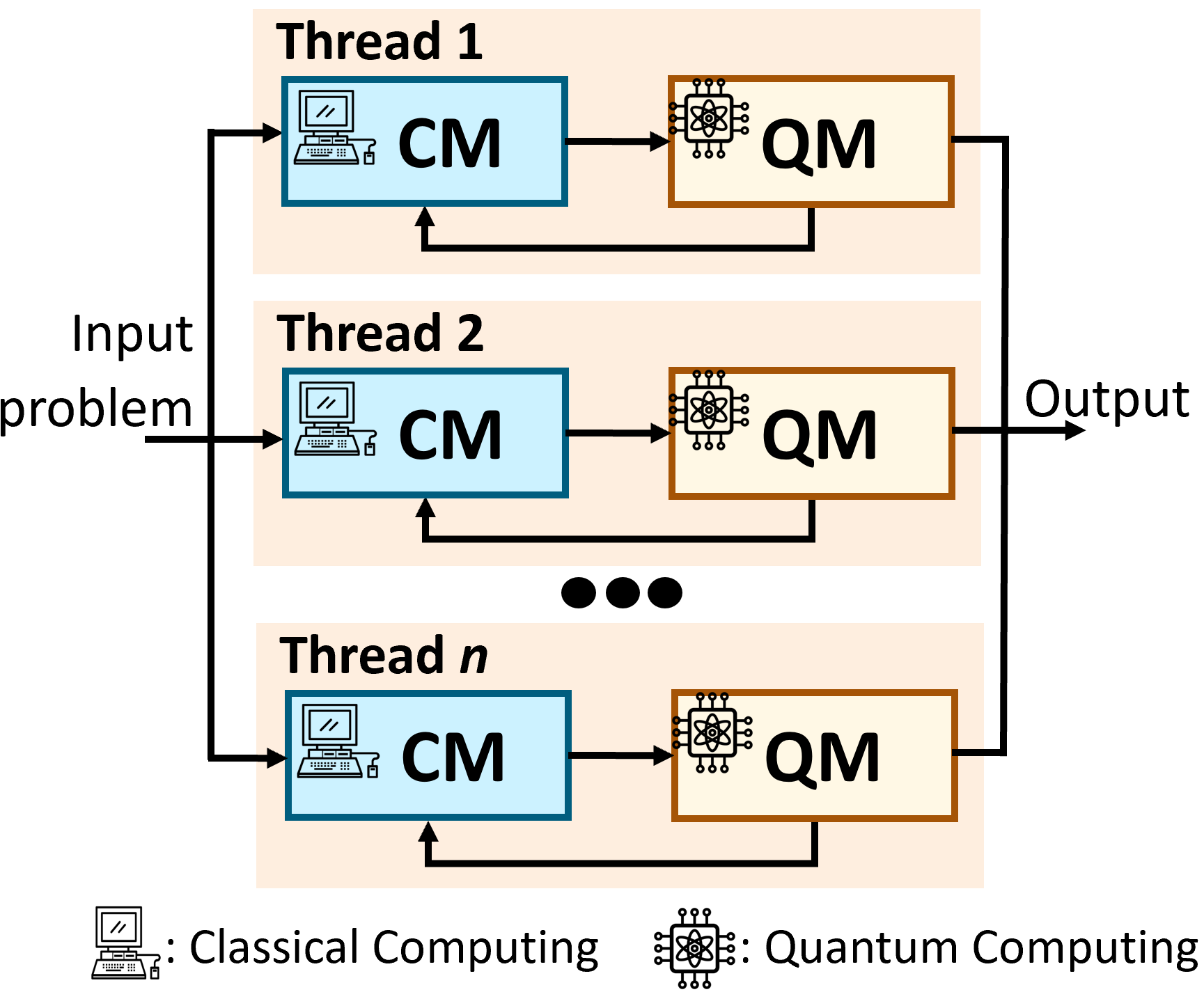}
    \caption{D-Wave's HSS schemes. CM = Classical Module. QM = Quantum Module.}
    \label{fig:hss}
\end{figure}

All solvers in HSS share the same structure, as illustrated in Fig. \ref{fig:hss}. In a nutshell, these algorithms are divided into two phases. First, the input problem is read and a group of parallel hybrid threads is created. Every thread includes a Classical Heuristic Module (CM), responsible for exploring the entire solution space, and a Quantum Module (QM), which, according to D-Wave "\textit{formulates quantum queries that are sent to a back-end Advantage QPU. Replies from the QPU are used to guide the heuristic module toward more promising areas of the search space or to find improvements to existing solutions}" \cite{leapCQM}. Then, all branches are independently executed and the best solution identified from the pool of threads is returned to the user.

\begin{figure*}[t!]
    \centering
    \subfigure[UC1\_12.]{\includegraphics[width=0.32\linewidth]{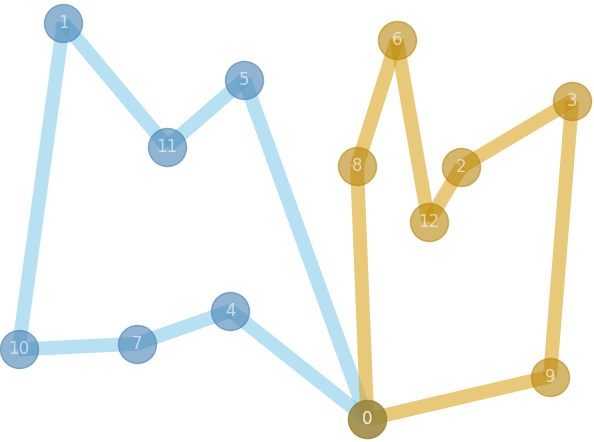}}
	\subfigure[UC1\_16.]{\includegraphics[width=0.32\linewidth]{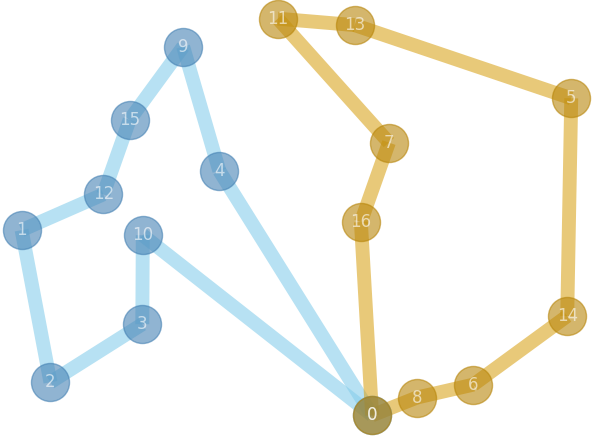}}
    \subfigure[UC1\_22.]{\includegraphics[width=0.32\linewidth]{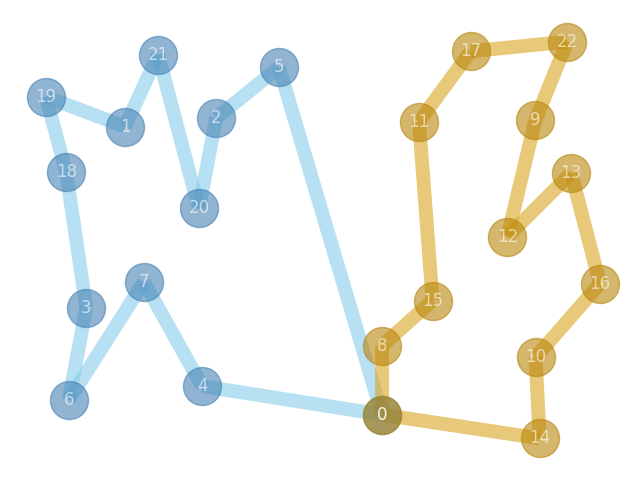}}
    \caption{Results obtained by \texttt{Q4RD} in the UC1.}
    \label{fig:UC1}
\end{figure*}

Following extensive laboratory research, partially described in \cite{osaba2024d}, it has been concluded that the method yielding the best results among the four implemented for UC1 and UC2 is the \texttt{NL-hybrid}. Therefore, the results presented for these cases have been obtained using the \texttt{NL-hybrid}. Nevertheless, for readers interested in the development aspects, we have publicly shared all the codes implemented for this research in \cite{PaperRep}.

Lastly, due to the intrinsic complexity of the mathematical formulation of UC3, the method employed for its implementation and resolution is the \texttt{CQM-Hybrid}, owing to its ease of use, efficiency, and flexibility. Further details are given in the upcoming Section \ref{sec:UC3}.

\section{Demonstration} \label{sec:exp}

This section is dedicated to demonstrating the applicability of the developed system across the three use cases examined. Each case has been tested on three distinct instances, consisting of 12, 16, and 22 visiting points, respectively. Each instance was generated specifically for this research, employing the nomenclature UC$X$\_$Y$ for its designation, where $X$ represents the use case number and $Y$ denotes the problem size. For the sake of transparency and replicability, the created benchmark and the obtained results are openly accessible in \cite{PaperRep}.

\subsection{UC1: Two drones and a single depot} \label{sec:UC1}

In Fig. \ref{fig:UC1}, we present the results obtained by \texttt{Q4RD} for the instances corresponding to UC1, where both drones commence and conclude their routes at the same depot. Notably, for comparative purposes, the quality of each route calculated by \texttt{Q4RD} has been assessed against a solution calculated by \texttt{Google} \texttt{OR-Tools}. For this purpose, an implementation openly shared has been used\footnote{https://developers.google.com/optimization/routing/tsp?hl=es-419}. In all examined cases, both methods yielded identical results, thereby demonstrating the reliability of the system developed in this paper.

\subsection{UC2: Two drones and two depots} \label{sec:UC2}

The primary distinction between UC2 and UC1 is the presence of two depots, each geographically situated at different locations. This configuration results in the creation of two completely independent routes, as the drones start and conclude their route in separate depots. That is, a drone is not permitted to begin its route at one depot and finish it at the other.

\begin{figure*}[t]
    \centering
    \subfigure[UC2\_12.]{\includegraphics[width=0.30\linewidth]{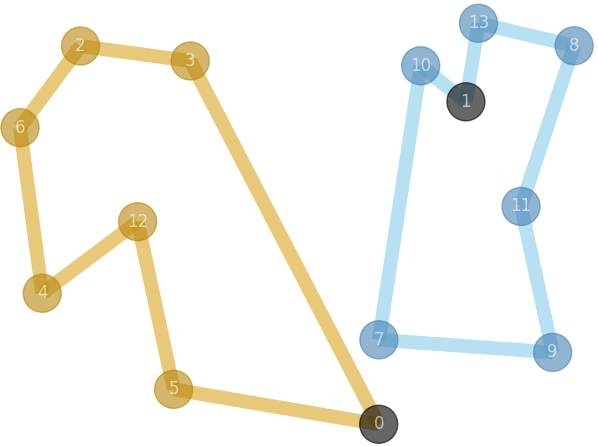}}
	\subfigure[UC2\_16.]{\includegraphics[width=0.30\linewidth]{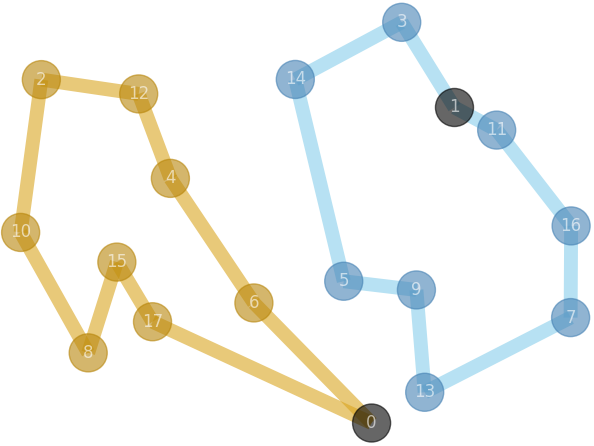}}
    \subfigure[UC2\_22.]{\includegraphics[width=0.30\linewidth]{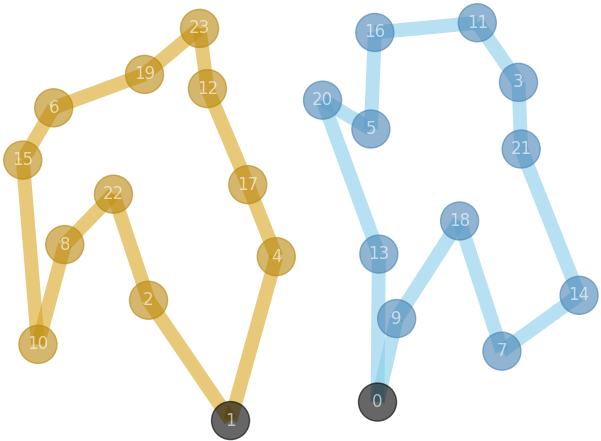}}
    \caption{Results obtained by \texttt{Q4RD} in the UC2. The depots are colored black.}
    \label{fig:UC2}
\end{figure*}

Consequently, once the \textit{Clustering phase} is completed and the points that will constitute each route are established, the subsequent step is to assign a depot to each route. This assignment is carried out as follows: first, the geometric center is computed for each subgraph:
\begin{equation}\label{eq:center}
x_c = \frac{1}{n} \sum_{i=1}^{n} x_i, \quad y_c = \frac{1}{n} \sum_{i=1}^{n} y_i
\end{equation}
After that, the distances from each depot to the centroid of each cluster are measured. The final assignment is the one that results in the shortest total distance, ensuring the depots are assigned in a way that minimizes the overall travel distance to the subgraphs.
We depict in Fig. \ref{fig:UC2} the results obtained by \texttt{Q4RD} for the instances corresponding to UC2.

\begin{figure*}[t]
    \centering
    \subfigure[UC3\_12.]{\includegraphics[width=0.30\linewidth]{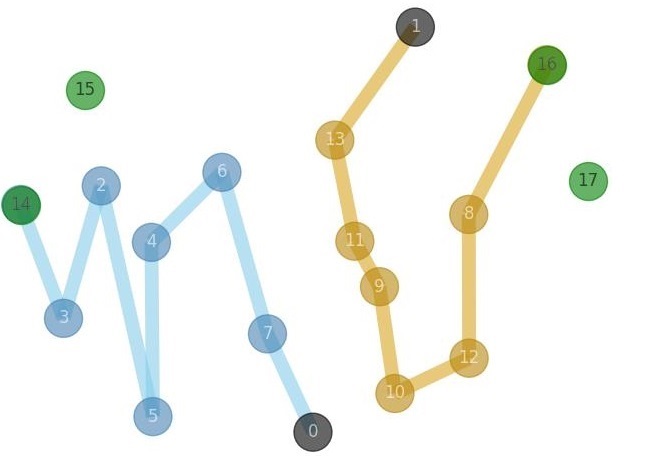}}
	\subfigure[UC3\_16.]{\includegraphics[width=0.30\linewidth]{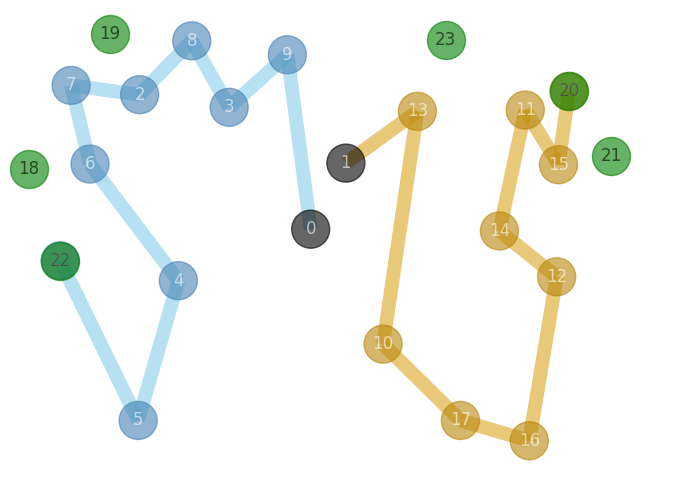}}
    \subfigure[UC3\_22.]{\includegraphics[width=0.30\linewidth]{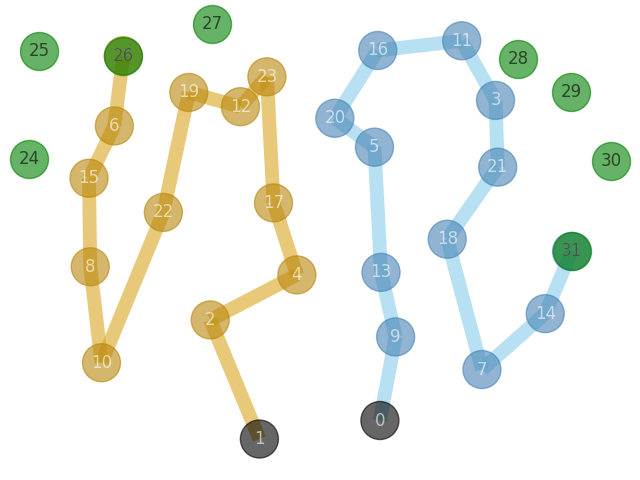}}
    \caption{Results obtained by \texttt{Q4RD} in the UC3. The depots are colored black.}
    \label{fig:UC3}
\end{figure*}

\subsection{UC3: Two drones, two depots and itinerant charging stations} \label{sec:UC3}

The main goal of the routing problem addressed in this UC3 is to design an \textbf{open} route that minimizes the total cost. This route must start from one of the depots, visit each mandatory point, and conclude at one of the multiple charging points distributed across the scenario. It is important to emphasize that the number $M$ of charging stations utilized in each instance is equal to $N$/3, where $N$ represents the number of the visiting points, as previously mentioned. These stations have been randomly distributed in what could be considered the periphery of the mandatory visiting nodes. This aspect becomes much clearer upon a brief examination of Fig. \ref{fig:UC3}. Furthermore, it should be clarified that these charging stations are not considered in the first step of the solving process, that is, the \textit{Clustering phase}.

Given the necessity of considering charging stations and concluding the route at just one of them (discarding the rest), we have crafted an optimization problem tailored to this specific use case. Thus, we have developed a mathematical formulation suitable on D-Wave's CQM model, enabling the problem to be solved using the above mentioned \texttt{CQM-Hybrid} approach. For this paper to be self-contained, we proceed now with the mathematical description of the problem designed.

\textbf{Codification and variables}: A binary encoding referred to as \textit{node-based} \cite{osaba2024solving}, has been employed to represent both the depot and the mandatory visiting points. Consequently, a set $X=\{X_0,\dots,X_n\}$ of lists has been established, where each $X_i$ corresponds to a single node $i$, with $n$ being the total number of visiting nodes and \textit{0} denoting the depot. Moreover, $X_i=\{x_{i,0},\dots,x_{i,n}\}$, where $x_{i,p}$ is a binary variable indicating the position of node $i$ along the route. Specifically, $x_{i,p}$ is 1 if node $i$ is visited in position $p$ of the route, and 0 otherwise.

Additionally, a set of binary variables $Y=\{y_1,\dots,y_m\}$ is introduced, where $m$ denotes the total number of charging points. In this set, $y_{i}$ is 1 if the route ends at the charging point $i$, and 0 otherwise.

\textbf{Objective}: The problem has a single objective, which is to minimize the total cost of the constructed route.

\begin{equation}
\begin{split}
f(x) = & \sum_{i=1}^{n} c_{0 i} x_{i 1} + \sum_{p=1}^{n-1} \sum_{\substack{i, j = 1 \\ i \neq j}}^{n} c_{i j} x_{i p} x_{j p+1} \\
& + \sum_{i=1}^{n} \sum_{j=1}^{m} c_{i j+n} x_{i n} y_{j}
\end{split}
\end{equation}

\textbf{Problem constraints}: The aforementioned objective is constrained by three distinct restrictions, which are:

\begin{itemize}
    \item \textit{Visiting Nodes consistency}: a visiting node $i$ must be visited exactly once. 

    \begin{equation}
    \sum_{p=1}^{n} x_{i p} = 1, \quad \forall i \in \{1, \dots, n\}
    \end{equation}
    
    \item \textit{Location consistency}: Only one visiting node $i$ can be assigned to each position $p$ in the route. In other words, two nodes cannot be visited at the same time.

    \begin{equation}
    \sum_{i=1}^{n} x_{i p} = 1, \quad \forall p \in \{1, \dots, n\}
    \end{equation}
    
    \item \textit{Charging point consistency}: The number of charging points to visit must be exactly one

    \begin{equation}
    \sum_{i=1}^{m} y_{i}=1
    \end{equation}
    
\end{itemize}

Finally, to ensure that every route starts from the depot, we establish $x_{0,0}=1$. We show in Fig. \ref{fig:UC3} the solutions got by \texttt{Q4RD} for the instances corresponding to UC3. Due to the unique characteristics of this use case, it has not been possible to conduct a comparison with a classical method, as done for the previous use cases. This task is designated as future work.

\section{Conclusions \& Further Work} \label{sec:conc}

In this paper, we have presented a hybrid approach, coined \texttt{Q4DR}, to solving drone routing problems by leveraging the capabilities of quantum computing. Being one of its main contributions, our approach combines quantum annealing and gate-based quantum computing techniques, employing D-Wave's hardware and the \texttt{Eclipse Qrisp} programming language. The proposed system effectively addresses designed drone routing problems, demonstrating its utility through three use cases of increasing complexity. 

Despite the encouraging results, several challenges persist. The current constraints of quantum hardware, including the limited number of qubits and issues related to decoherence and noise, limit the size and complexity of problems that can be efficiently addressed. Furthermore, the integration of classical and quantum computing paradigms (including the combination of annealing and gate-based computing) needs further refinement to fully leverage the strengths of hybrid methods.

Numerous exciting challenges and opportunities have been identified for future endeavors:

\begin{itemize}
    \item With respect to the \textbf{problem definition}:
    \begin{itemize}
        \item An extension of \texttt{Q4DR} to deal with a broader range of restrictions and objectives, making it applicable to a wider variety of scenarios. Potential new features might include priority nodes or accounting for drone capacities.
        \item Consider other types of unmanned vehicles in order to have an heterogeneous fleet of units.
    \end{itemize}

    \item Regarding \textbf{technology and the solving approach}:
    \begin{itemize}
        \item Employ complexity reduction mechanisms to efficiently manage larger scenarios by minimizing the problem's size while preserving all constraints.
        \item Explore the use of more advanced quantum algorithms and hybrid techniques to improve solution quality and scalability.
        \item Investigate the potential of emerging quantum hardware technologies to overcome current limitations and enable the solving of larger and more complex problems. 
        \item Conduct an extensive benchmarking using classical optimization methods to further validate \texttt{Q4DR}.
    \end{itemize}    
\end{itemize} 

By tackling these challenges and exploring new paths for enhancement, we aim to advance the field of quantum optimization and contribute to the practical application of quantum computing in logistics and route planning.

\bibliographystyle{IEEEtran}
\bibliography{IEEEexample}
\end{document}